# Hybrid iterating-averaging low photon budget Gabor holographic microscopy

Mikołaj Rogalski[1,*], Piotr Arcab[1], Emilia Wdowiak[1], José Ángel Picazo-Bueno[2,3], Vicente Micó[2], Michał Józwik[1], Maciej Trusiak[1,^]

[1]Institute of Micromechanics and Photonics, Warsaw University of Technology, 8 Sw. A. Boboli St., 02-525 Warsaw, Poland

[2]Departamento de Óptica y Optometría y Ciencias de la Visión, Universidad de Valencia, C/Doctor Moliner 50, 46100 Burjassot, Spain

[3]Biomedical Technology Center, University of Muenster, Mendelstr. 17 D-48149, Muenster, Germany

*mikolaj.rogalski.dokt@pw.edu.pl

^maciej.trusiak@pw.edu.pl

## Abstract

One of the primary challenges in live cell culture observation is achieving high-contrast imaging with minimal impact on sample behavior. Quantitative phase imaging (QPI) techniques address this by providing label-free, high-contrast images of transparent samples. Measurement system's influence may be further reduced by imaging samples under low illumination intensity (low photon budget – LPB), thereby minimizing photostimulation, phototoxicity, and photodamage, and enabling high-speed imaging. LPB imaging is particularly challenging in QPI due to significant camera shot noise and quantification noise. Digital in-line holographic microscopy (DIHM), working with or without lenses, is a QPI technique known for its robustness to quantification noise. However, reducing simultaneously the camera shot noise and the inherent in-line holographic twin image disturbances remain a critical, yet unaddressed, challenge. This study introduces an innovative iterative Gabor averaging (IGA) algorithm designed specifically for filling this important scientific gap in multi-frame DIHM under LPB conditions. We evaluated the performance of the IGA on simulated data showing that it outperformed traditional algorithms in terms of reconstruction accuracy under high noise conditions. Those results were corroborated by experimental validation involving high-speed imaging of dynamic sperm cells and a phase test target under significantly reduced illumination power. Additionally, the IGA algorithm proved successful in reconstructing optically thin samples, which typically produce low signal-to-noise ratio holograms even under high photon budget conditions. These advancements facilitate photostimulation-free and high-speed imaging of dynamic biological samples and enhance the capability to image samples with extremely low optical thickness, potentially transforming various applications in biomedical and environmental imaging.

# Introduction

Imaging live bio-samples, such as cell cultures, poses a significant challenge in optical microscopy due to their inherent transparency, which leads to generally very low contrast. Quantitative phase imaging (QPI) techniques have emerged as a popular solution for visualizing transparent objects with high contrast, independent of their absorption properties [1], [2]. QPI methods are based on measuring the phase delay between the light wave that passes through the sample and the light wave that passes through its surrounding medium. As long as the optical thickness of the sample and the surrounding medium differs, QPI methods can provide high-contrast images even for completely transparent samples. Additionally, QPI reconstruction yields quantitative information about the sample's optical thickness [3].

Among the various QPI methods, digital in-line holographic microscopy (DIHM) is particularly noteworthy [4]. DIHM setups distinguish themselves from other QPI techniques due to their simplicity and cost-effectiveness. For example, DIHM can be implemented in classical brightfield microscopes with added coherent illumination module [5], [6]. But probably the biggest advantages come in lensless microscopy configuration (lensless DIHM) [7], [8], which, aside from being extremely simple, size-optimized and cost-effective (as it eliminates the need for microscope objectives and embodiments), offers exceptionally large fields of view (easily over 100 mm$^2$) with reasonable spatial resolution (typically around 1-2 µm) – a so-called high space-bandwidth-product imaging [9]. By eliminating microscope objectives, lensless phase microscopes also avoid optical aberrations, depth of field, and working distance limitations.

DIHM belongs to a broader category of computational microscopy techniques [10] where data captured by the camera requires appropriate processing to reconstruct physically meaningful information. Although DIHM reconstruction ideally requires only a single image (in-line hologram), single-frame reconstructions are plagued by the twin image effect [7], [8] – a disturbance numerical phenomenon resulting from lack of phase information at the camera plane. Minimizing this effect – and thus improving QPI results – from a single hologram is an ill-posed problem, which can be partially addressed through regularization [11], [12] or deep learning approaches [13], [14]. More robust methods for twin image minimization involve capturing multiple images under different imaging conditions (e.g., different camera positions along the optical axis [15], [16] or different illumination wavelengths [17]–[19]). Multi-frame reconstruction is a well-posed problem that can efficiently converge to the true solution. However, its main drawback is the need to collect multiple images, which is usually done sequentially, reducing temporal resolution and increasing hardware burden.

Another significant challenge in imaging live cell cultures is to avoid the photostimulation [20] and in more severe cases phototoxicity [21], [22] and photodamage [23], [24] of the cells due to the influence of illumination light. This influence can be mitigated by reducing the illumination radiation dose, leading to low photon budget (LPB) imaging conditions characterized by low illumination power relative to the measurement time (camera exposure time). LPB conditions can also arise when imaging samples in high speed scenario [25] or using exotic wavelengths that greatly fall outside the detector's optimal quantum efficiency range [26], [27].

LPB imaging conditions in QPI [28]–[32] are challenging due to two primary factors: (1) low signal-to-noise ratio (SNR), resulting in images marred by camera shot noise, and (2) low signal intensity, leading to quantification noise (when the signal is sampled by a low number of camera gray levels). Interestingly, as shown by our previous results [33], DIHM demonstrates great potential in LPB imaging using single-frame setups due to its reduced sensitivity to quantification noise compared to other QPI systems. This capability allowed for high-accuracy DIHM reconstructions even from images collected with only few camera's gray levels. Nonetheless, the persistent challenges include the presence of camera shot noise and twin image perturbation, which are very difficult to minimize simultaneously from a single acquired image (although training deep neural network with noisy data [28]–[31], [34] or employing appropriate regularizations [12], [35], [36] may potentially allow for enhanced reconstructions).

Intuitively, the drawbacks of single-frame DIHM methods in LPB imaging should be mitigated by multi-frame DIHM techniques. As previously stated, multi-frame iterative DIHM algorithms are more robust against twin image disturbance than single-frame methods. Additionally, acquiring multiple images can also aid in averaging shot noise. However, the conventional iterative QPI methods are generally known for their low accuracy when working with noisy data [37]–[39], and as shown by our novel results presented in this paper, this is also the case in DIHM.

In this work, we propose a novel, open-source [40], multi-frame DIHM reconstruction framework specifically designed to simultaneously minimize shot noise and twin image disturbance in LPB conditions. We evaluate the robustness of the proposed method through numerical simulations, demonstrating, for the first time to the best of our knowledge, the ability to effectively reduce both types of noise without compromising image resolution. The numerical results are validated by experimental LPB imaging outcomes of (1) a static two-photon polymerization (TPP) printed phase object measured in a lensless DIHM setup and (2) a dynamic sperm cell sample imaged in a high-speed brightfield microscope DIHM. Furthermore, we show that the proposed algorithm can be successfully applied to measure extremely low phase change samples by imaging a phase test target with a thickness of only 16 nm (0.08 rad phase thickness) – a very challenging case even under high photon budget (HPB) imaging conditions.

# Methods

## DIHM and Lensless DIHM systems

Figure 1(a) illustrates the scheme of the DIHM system used in this study that, essentially, was the same one as presented in Ref. [5], [6]. The system consisted of a regular upright microscope embodiment (Olympus BX-60 with UMPlanFl 20X 0.46NA objective) where a fiber coupled diode laser was externally inserted to provide coherent illumination in the system. The multiwavelength laser source (Blue Sky Research (SpectraTec 4 STEC4 405/450/532/635 nm) provided RGB (450/532/635 nm) illumination and a RGB camera (Ximea USB3 MQ042CG-CM, CMOS sensor type, 2048 × 2048 pixels, 5.5 μm pixel pitch, 90 fps) was used to record the RGB holograms in a single snapshot. As in previous setups [5], [6], the sample was moved away from the microscope lens so that the image plane was located before the digital sensor plane. Under these conditions, the camera recorded three color-coded in-line holograms with free-space propagation defocus each one of them from the image plane as in a classical Gabor configuration. LPB imaging conditions were introduced by lowering the camera exposure time, extremely important regime in high-speed imaging scenarios.

Figure 1(b) depicts the scheme of the lensless DIHM system employed. The laser used in this system (CNI Lasers MGL-FN-561-20 mW, λ = 561 nm) emitted light with linear polarization. To facilitate easy control of the illumination power, a linear polarizer was positioned directly after the laser. By rotating the polarizer, the illumination intensity could be easily adjusted, ranging from zero to the maximum laser power. The light was then coupled into a fiber (Thorlabs P1-460B-FC-

1), with the fiber end aligned on-axis with the camera (ALVIUM Camera 1800 U-2050 m mono Bareboard, pixel size 2.4 × 2.4 μm, 5496 × 3672 pixels), at a distance of approximately 20 cm from the camera plane. The camera was mounted on a translational stage (Thorlabs KMTS25E/M), allowing for precise control over the distance between camera and sample planes.

Despite the significant differences, both systems are based on the same physical principles (Gabor in-line holography), allowing for similar processing paths to reconstruct the data acquired. In both systems, the sample was illuminated with a quasi-monochromatic wave illumination. The sample's complex optical field was then propagated in free space to the camera (in the lensless DIHM system) or to the objective focal plane and subsequently imaged onto the camera (in the DIHM system). The camera captures the amplitude part of the free-space propagated (defocused) optical field (in-line hologram), which can be then numerically backpropagated [41], [42] to the sample/image plane to retrieve its complex (amplitude and phase) distribution – this is known as the Gabor holography reconstruction scheme (GHR) [43].

Due to the lack of the phase information in the recording process performed at the camera plane, GHR is compromised by the twin image presence. As discussed in the Introduction, this effect can be minimized via iterative approach introducing data multiplexity – collecting several different holograms using varied illumination wavelengths or defocus distances. In the used DIHM system, data multiplexity was achieved by illuminating the sample with three different RGB wavelengths, thereby producing three different holograms simultaneously, each captured by a different color channel of the RGB camera. In the lensless DIHM system, data multiplexity was achieved by moving the camera along the optical axis and collecting sequentially three holograms with different defocus (sample-to-camera) distances.

Used systems differ in their imaging capabilities. The DIHM system's imaging resolution is diffraction-limited (to 0.8 μm), with a field of view constrained by the camera sensor dimensions and the optical magnification (to 0.39 x 0.39 mm), and its temporal resolution is limited by the camera framerate (90 fps). The lensless DIHM system's resolution is limited by the camera pixel size (2.4 μm), its field of view is limited solely by the camera sensor dimensions (13.19 x 8.81 mm), and its temporal resolution is restricted by the need to scan the camera along the optical axis and acquire multiple images (a single unoptimized measurement of three scanned heights took approximately 2 seconds).

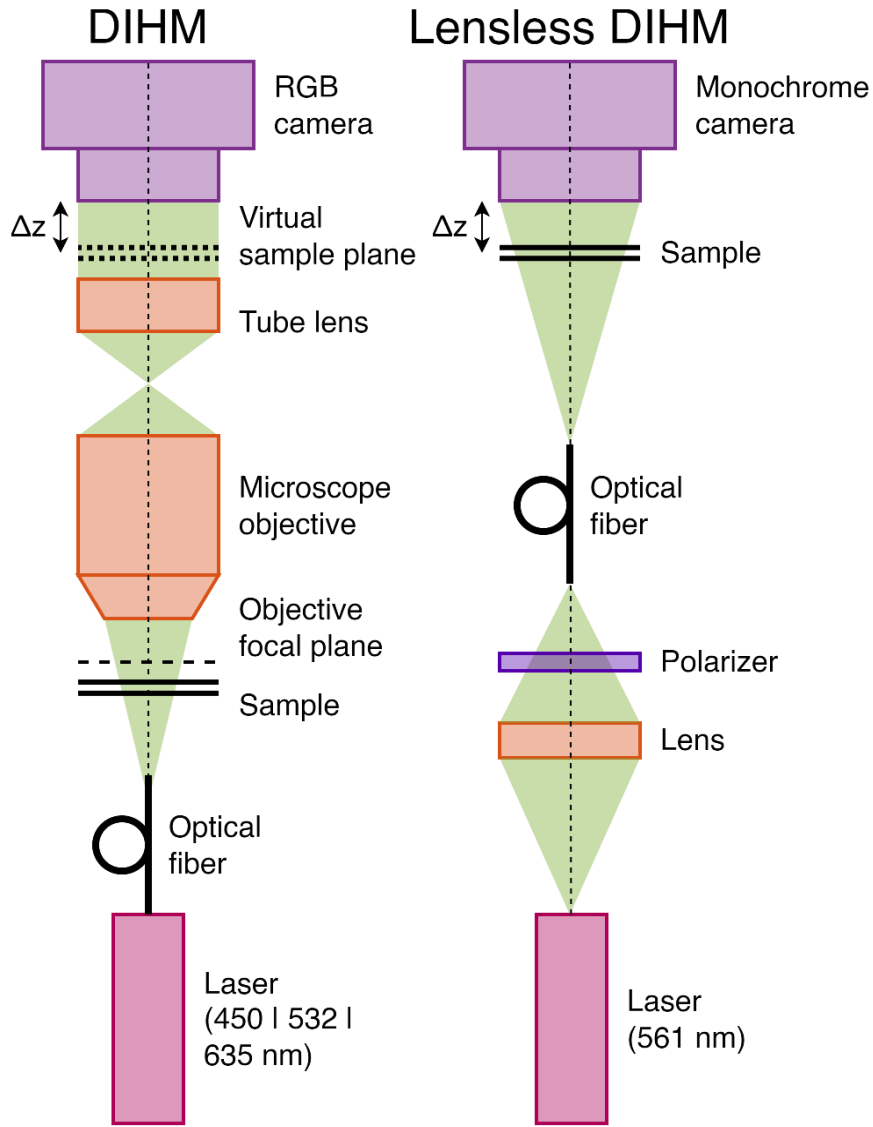

*Fig. 1. Scheme of (left) lens-based and (right) lensless DIHM systems.*

## Proposed iterative Gabor averaging algorithm

Traditionally, twin image disturbance can be minimized from multiplexed holograms using various iterative algorithms, such as Gerchberg-Saxton (GS) [44], hybrid input-output [45], or conjugate gradient [46]. These methods rely on the iterative numerical propagation of the estimated optical field between different hologram planes, updating the optical field in each plane based on the collected hologram. This approach converges to the true solution if the acquired holograms accurately represent the amplitude part of the optical field. However, this assumption does not hold in the presence of noisy holograms, which also contain information unrelated to the reconstructed optical field (e.g., camera shot noise).

Figure 2 presents a simple simulation showing the performance of iterative GS algorithm in terms of twin image minimization for noisy and noise-free DIHM data. In this simulation, the GS algorithm was provided with two multiplexed holograms (collected with different defocus distances) of a purely amplitude object and performed for 100 iterations. For shot noise-free data, the GS converged to true solution, efficiently minimizing the twin image contribution. However, for shot noise-spoiled data, the reconstruction root mean square (RMS) error increases with the number of iterations. Closer analysis of the reconstruction results reveals that GS managed to minimize the twin image disturbance at the cost of shot noise multiplication, which is extremely alarming effect in LPB imaging. To the best of our knowledge, this property was not shown in the DIHM before, although it is consistent with observations made for iterative phase retrieval methods in different QPI systems [37]–[39].

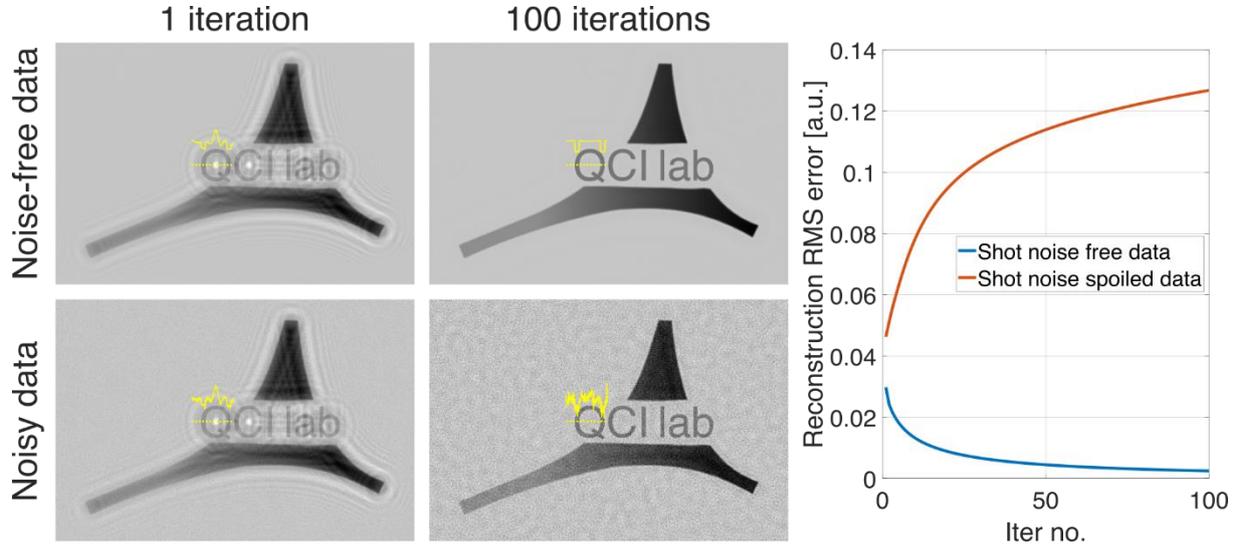

*Fig. 2. Simulation results of GS twin image minimization for shot noise-free and shot noise-spoiled data. There were simulated holograms (collected with different defocus) of a purely amplitude object with amplitude values ranging from 0 to 1 [a.u.]. Noisy data holograms were spoiled with a Gaussian noise of 0.05 standard deviation.*

When having multiple images of the same object, which differ only by the shot noise (of similar level in each image), the most straightforward and natural way to minimize the noise is to average the acquired images [47]. Shot noise can also be minimized in single-frame scenarios [48] using methods such as spatial filtering [49], iterative regularization [35], frequency filtering [50] or deep learning [51]. However, for successful denoising, single-frame methods: (1) require some 'a priori' knowledge about the measured object and/or noise characteristics, (2) may incur in high computational costs, and (3) may reduce spatial resolution (blurring the image). Single-frame denoising methods can also be applied in multi-frame scenarios, either by first averaging the acquired images and then performing single-frame denoising or by starting with single-frame methods applied to each image and then averaging the results.

In the case of multi-frame shot noise minimization of multiplexed DIHM data, frame averaging cannot be directly applied to the collected images, as the input DIHM holograms contain different object information (acquired with different defocus/illumination wavelengths). However, the averaging can be applied to the reconstructed holograms (numerically backpropagated to the object plane) – we denote this operation as Gabor averaging (GA) method. Each reconstructed hologram contains the same in-focus object information and differs by the presence of shot noise and partially by twin image disturbance (especially higher spatial frequency twin image information). Therefore, averaging them allows for not only shot noise reduction but also partial twin image presence reduction, though not to the accuracy of iterative DIHM algorithms.

Depending on the shot noise level, different multi-frame reconstruction approaches may be more successful. For low shot noise levels, the twin image issue is the main source of noise, favoring iterative DIHM methods. For strong shot noise, its influence is dominant, favoring the GA approach. However, in mixed scenarios where both noise types are similarly significant, neither method will provide accurate denoising results. This is precisely the gap in the state-of-the-art we address in this contribution.

Here, we present a novel algorithm called iterative Gabor averaging (IGA), which combines the advantages of iterative and GA approaches to provide a multi-frame LPB DIHM reconstruction with both noise types minimized. To design an accurate algorithm, the properties of both noise types should be analyzed. Figure 3(a) presents a numerically simulated amplitude-type object, and Fig. 3(c) shows its reconstruction spoiled by shot noise, twin image disturbance, and both noise types simultaneously. As observed, the noise types are visually distinct: shot noise is uniformly distributed across the entire reconstructed image, while twin image disturbance manifests as a diffraction pattern overlaying the measured object. The difference between the noise types can be distinguished in the frequency domain. Figure 3(b) presents a plot of the horizontal spatial frequencies of the noise error maps (calculated as the simulated object minus the noise-spoiled reconstruction). Shot noise is uniformly distributed across different spatial frequencies, while twin image noise is primarily concentrated in low spatial frequencies.

The proposed IGA method leverages the above-mentioned noise properties by combining iterative GS phase retrieval reconstruction for low-frequency information (where twin image disturbance is dominant) with GA reconstruction for high-frequency information (where shot noise is dominant). The IGA processing path is presented in Fig. 3(d). The algorithm takes as input several (at least two) in-line holograms collected with different defocus distances/illumination wavelengths, which are then processed by two algorithm paths independently. In the first patch, each input hologram is low-pass filtered with a Gaussian kernel of user-chosen standard deviation ($\sigma$). The low-pass filtered holograms are then provided to the GS phase retrieval algorithm, which, thanks to the shot noise minimization via low-pass filtering, retrieve the twin image-free (but blurred) reconstruction without significant noise amplification. Simultaneously, in the second path, the GA method is applied to obtain shot noise averaged reconstruction from the input data. This reconstruction is then high-pass filtered with the same Gaussian kernel (same $\sigma$) as used for low-pass filtering in the first path. Finally, the results from both paths are combined by summation giving twin image minimized, shot noise averaged reconstruction without information blur (see Supplementary Document 1 sections 1-4 and the released open-source codes [40] for the implementation details).

The proposed algorithm is based on the concept of combining high-pass and low-pass image filtering, which is fundamental in many image processing applications [52]–[54]. Notably, the proposed method is relatively simple, resulting in low computational costs (insignificantly larger than the combined time of the GS and GA methods). The simplicity of the method also enhances its robustness, as it does not involve many steps that could introduce algorithmic errors. The algorithm's outcome depends on two user-defined parameters: (1) the number of iterations of the GS algorithm ($T$), and (2) the Gaussian kernel standard deviation $\sigma$. Increasing $T$ enhances twin image minimization accuracy but also extends reconstruction time. The $\sigma$ value balances the influence between the GA and GS algorithm components – the larger the $\sigma$, the stronger the shot noise minimization and the weaker the twin image minimization. The proposed novel open-ended numerical framework for LPB holographic denoising can be easily modified while maintaining its core novel idea of hybrid GA and iterative DIHM processing. For example, the GS algorithm can be replaced with a different iterative phase retrieval method [45], [46], or the Gaussian kernel high-pass/low-pass filtering can be substituted with another filtering method [50]. Especially, algorithm's performance could be further improved by adjusting the high-pass/low-pass filtration to the system shot noise characteristics [55], [56] (also resulting in algorithm automatization by eliminating the need to manually set σ parameter).

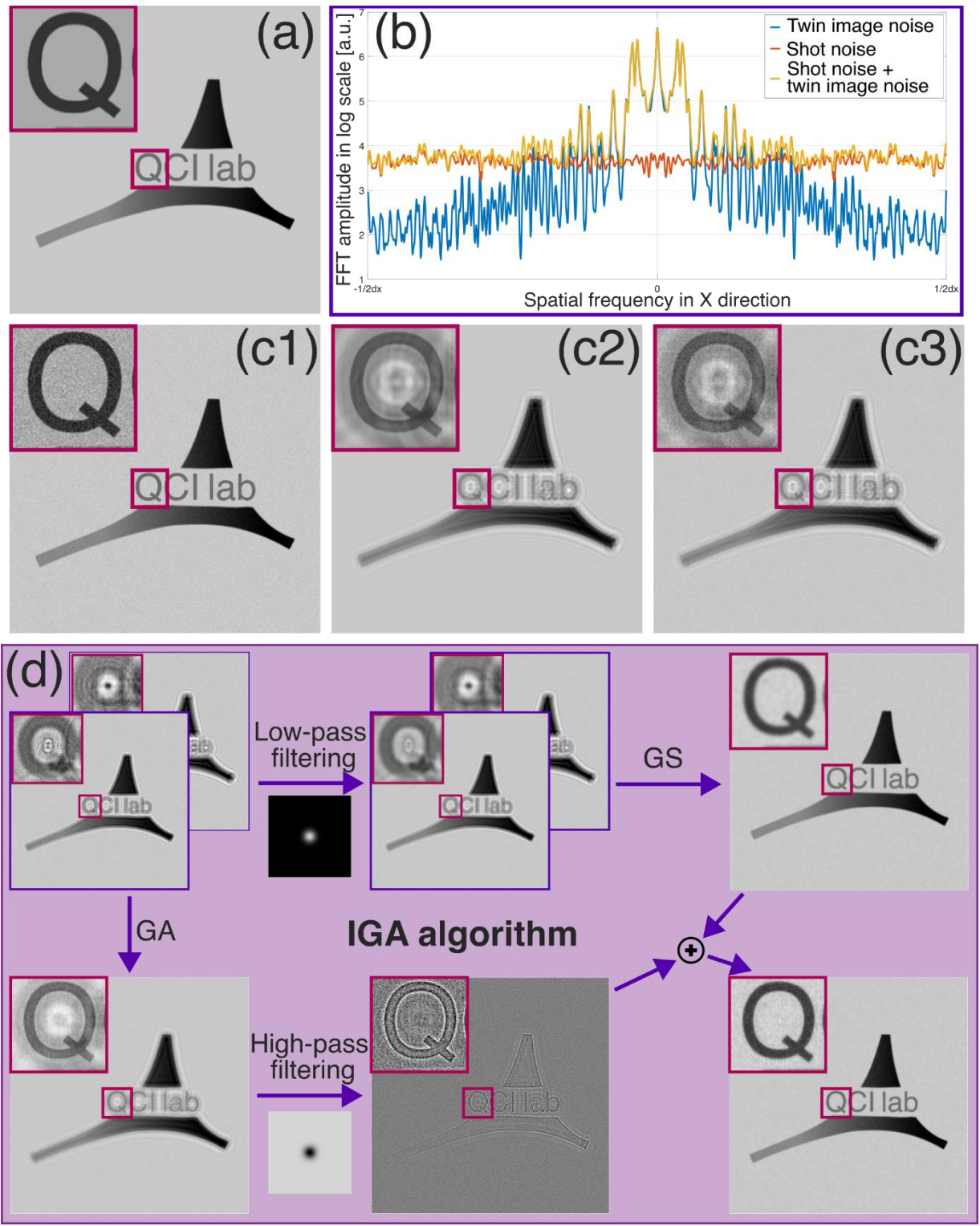

*Fig. 3. (a) Simulated purely amplitude object and (c1) object spoiled with shot noise, (c2) twin image noise and (c3) both noise types. (b) Fourier transform (horizontal frequencies) of noise error maps (differences between simulated object and the noise spoiled image) from (c), showing the frequency response of different noises. (d) Proposed IGA algorithm processing path for reconstruction of the object shown in (a) from two in-line holograms collected with different defocus distance and spoiled with shot noise (shown in top-left corner of (d)). In order to make the algorithm scheme clearer only the amplitude part of reconstructed optical fields are shown for GS, GA and IGA methods.*

# Results

## Simulations

To numerically evaluate the proposed algorithm, we simulated holograms acquired using the lensless DIHM system under varying shot noise levels. In the simulated system, three holograms were acquired at different camera-sample distances (2, 3, and 4 mm). The camera pixel size was set to 2 µm, and a perfectly coherent light source of 500 nm wavelength was simulated. The measured object was modeled as a purely phase object (our lab's logo) with phase values ranging from -0.4 to 0.6 rad, as shown in Fig. 4(a). Camera shot noise was simulated as Gaussian distribution noise with standard deviation (std) varying from 0 to 0.3 [a.u.], resulting in a minimal hologram peak SNR (PSNR) of 10.46 dB. The simulated holograms were reconstructed using the GS, GA, and IGA ($\sigma$ = 1, 4) methods. The GS and IGA methods were performed with T=100 iterations to ensure good convergence of the final result. Additionally, classical GHR phase retrieval was performed using single-frame data (holograms collected at a single camera-sample distance of 2 mm).

Figure 4(b) shows the plot of phase RMS error in relation to the shot noise std and hologram PSNR. As observed, for low shot noise levels, the GS and IGA methods achieved similarly good results with lower RMS error than straightforward GA and GHR. For higher shot noise levels, the GS performance decreased significantly compared to the other investigated algorithms – we showcase this disadvantage of GS for the first time, to the best of our knowledge, and link it to detrimental "convergence" to dominant noise (results presented in Fig. 2). The proposed IGA method performed similarly well to GS in high SNR cases and similarly well to GA in low SNR cases, demonstrating that it feasibly merges the advantages of both methods without being affected by their drawbacks.

Both investigated $\sigma$ parameters resulted in comparable reconstruction accuracy. A larger $\sigma$ was slightly more suitable for higher shot noise, while a smaller $\sigma$ was slightly better for lower shot noise. Surprisingly, even a relatively small $\sigma$ parameter (with $\sigma$ = 1, where the GS part of the IGA method was dominant) allowed for significantly better performance than GS in shot noise spoiled data. This indicates that the $\sigma$ value is a fine-tuning parameter rather than a critical one that must be set carefully to enable high accuracy reconstruction. Basing on this conclusion, in further experiments we empirically set $\sigma$ equal 2 (between $\sigma$ = 1 and $\sigma$ = 4, which worked well for the entire range of investigated noise levels). Detailed discussion on influence of $\sigma$ parameter is present in Supplementary Document 1 section 5.

Figure 4(c) presents the exemplifying reconstructions (zoomed-in letter Q of the simulated object achieved by the investigated methods for the noise-free and noisy cases, while Fig. 4(d) shows a cross-section through the central part of the reconstructed letter Q for the noisy data shown in Fig. 4(c). It can be observed that GHR and GA reconstructions are spoiled by twin image disturbances, which manifest mainly as a false positive information in the center of the letter Q. The GS method minimizes the twin image almost perfectly, however, for noisy data, the shot noise is increased. The IGA outperforms the GA method in terms of twin image minimization, achieving results nearly as good as GS for shot noise-free cases. However, unlike GS, it also performs well for shot noise spoiled data, achieving similar shot noise levels to GA.

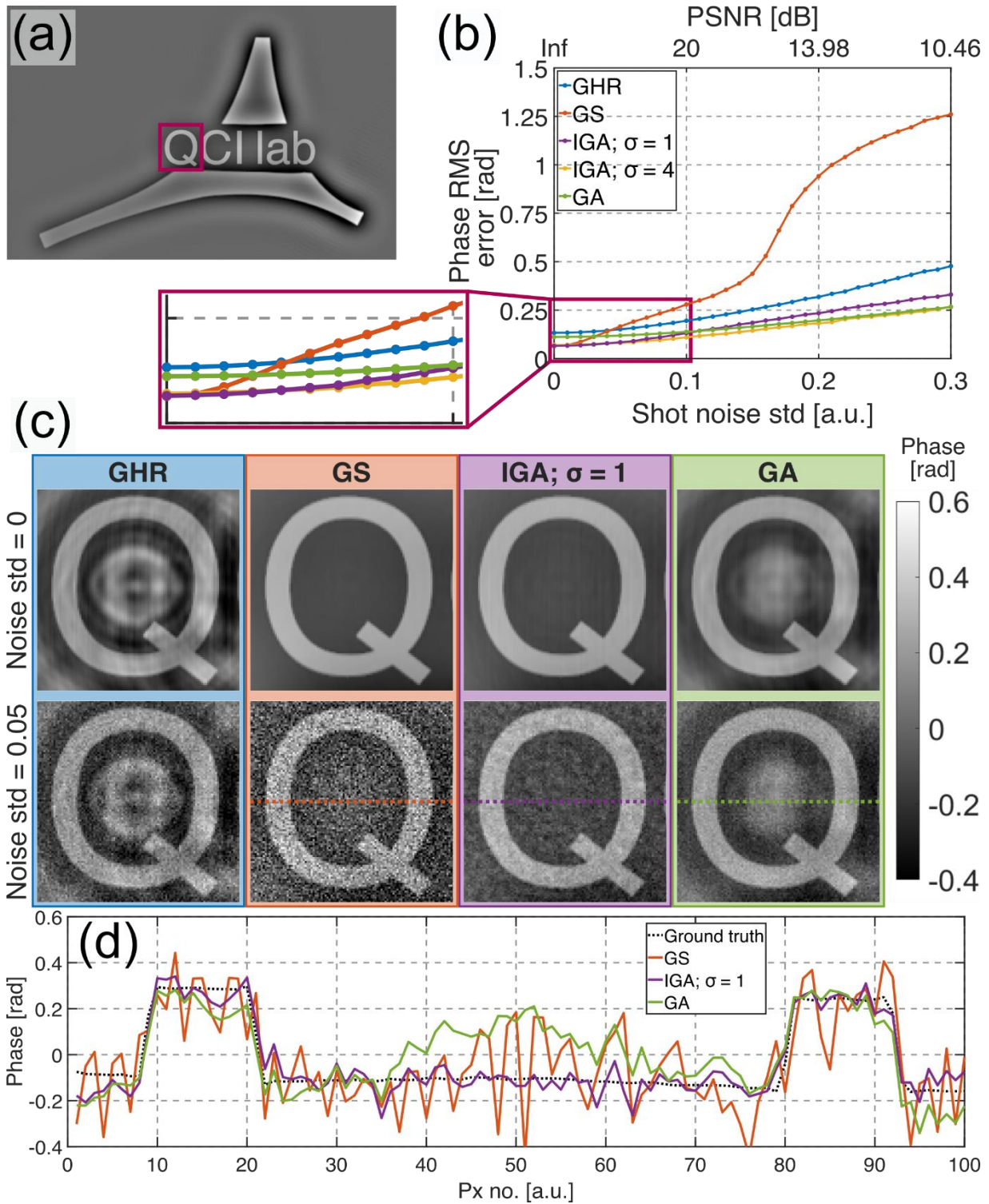

*Fig. 4. (a) Simulated purely phase object. (b) Reconstruction RMS errors for different algorithms and different input data noise levels. (c) Exemplifying reconstructions (zoomed-in letter Q from the investigated object) for shot noise free and shot noise spoiled data. (d) Cross-section for shot noise spoiled data shown in (c).*

## Experiments

Figure 5 presents the evaluation of a LPB QPI measurement in a lensless DIHM system. In the performed experiment, the measured object was a custom-made phase test target printed using the TPP technique (see Supplementary Document 1 section 6 for details). Three datasets (HPB, LPB1, and LPB2) were collected, each containing three in-line holograms recorded at different camera-sample distances (3, 4, and 5 mm). The HPB dataset was collected with laser illumination irradiance (L) of 2.71 µW/cm$^2$ (measured at the sample plane with Thorlabs S120C photodiode power sensor). For the LPB1 and LPB2 datasets, L was reduced to 0.21 and 0.11 µW/cm$^2$, respectively. The camera exposure time was set to 10 ms for each dataset. This combination of laser irradiance and camera exposure time resulted in the acquisition of reference "bright" holograms in the HPB data (utilizing most of the 8 bits camera's dynamic range) and "dark" holograms in the LPB data (utilizing only the few first camera gray levels). The brightness of the measured holograms was quantified as their mean value (M) ± their standard deviation (ΔM). The M and ΔM values (given in camera gray level (cgl) units), along with exemplary collected holograms and their histograms, are shown in the first two rows of Fig. 5.

The acquired datasets were then reconstructed using the GS, GA, and IGA algorithms. The obtained phase results are shown in the bottom three rows of Fig. 5. These results were also compared quantitatively by calculating the std of the object-free area (S) marked with a yellow rectangle in Fig. 5. The results achieved by the GA algorithm were again spoiled by twin image disturbance, visible as a diffraction pattern around the measured object. The GS and IGA methods reduced most of the twin image noise, resulting in a significantly more uniform background around the object in the case of HPB data, as quantitatively confirmed by the S measure. The GS results for LPB data were spoiled by shot noise, whereas the GA and IGA methods seemed to be much more robust to its negative influence.

Interestingly, all three methods managed to reconstruct the LPB1 dataset with only slightly worse accuracy (S measurement) than the HPB dataset, despite the fact that the LPB1 dataset was collected with over 12 times lower illumination power than the HPB dataset. This once again indicates the ability of DIHM to work effectively with LPB data [33].

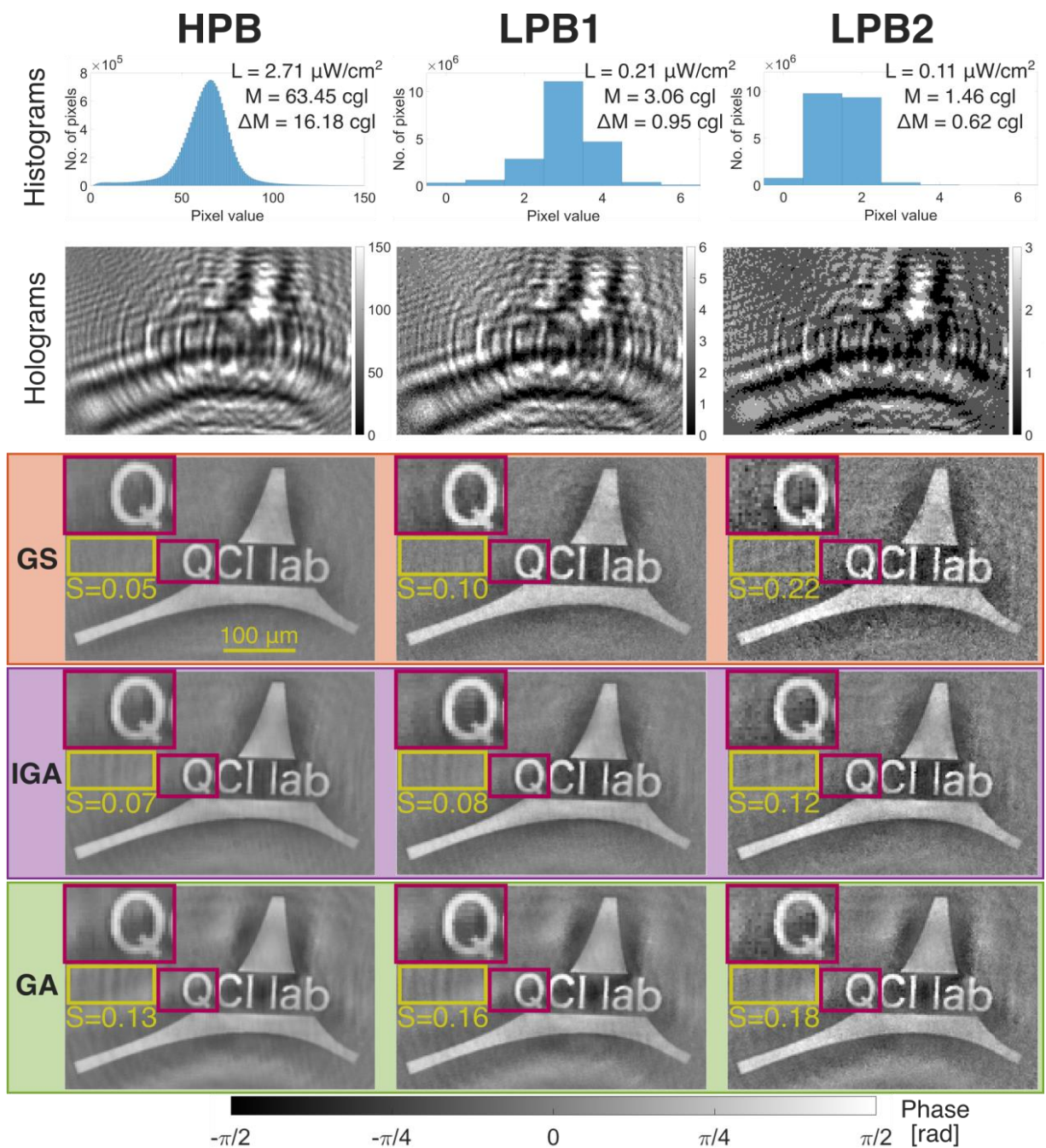

*Fig. 5. Top two rows – input in-line holograms and their histograms. Bottom three rows – GS, proposed IGA and GA reconstructions. Left, central and right column correspond to HPB, LPB1 and LPB2 datasets. L – laser irradiance measured in camera plane, M – hologram mean value, ΔM – hologram standard deviation, S – standard deviation of object-free area marked with yellow rectangle given in [rad] units.*

One of the potential applications where LPB imaging conditions are critical is high-speed imaging of dynamic biospecimens. In such cases, the low illumination radiation dose required to avoid stimulating or damaging the sample cannot be compensated by increasing the camera exposure time, as this would reduce both the measurement's temporal (as a direct consequence of the frame rate enlargement) and lateral (because higher exposure times will allow sample's movement thus averaging final acquired image and presenting blurred reconstructions) resolutions.

Typically, multi-frame DIHM systems are configured to collect the holograms needed for a single-instance reconstruction sequentially (as in lensless DIHM in Fig. 1), which limits the system's temporal resolution. However, in some configurations, the required input holograms can be collected simultaneously (e.g., using wavelength multiplexing, as in DIHM in Fig. 1), allowing for imaging at the camera's frame rate at the expenses of defining a higher effective pixel size.

Figure 6 presents the results of high-speed imaging of human spermatozoa (see Supplementary Document 1 section 7 for sample preparation and measurement details). In this experiment, we employed a wavelength multiplexing lens-based DIHM system (Fig. 1), collecting two sets of time-series data (HPB: $M \pm \Delta M$ = 74.05 ± 22.57 [cgl] and LPB: $M \pm \Delta M$ = 4.08 ± 1.81 [cgl]), acquired with different camera exposure times (HPB – 4 ms and LPB – 0.3 ms) but with the same illumination laser power. Each dataset comprised 300 RGB images collected at 90 fps. Each of the three image channels was treated as a single hologram collected with a different illumination wavelength. These three holograms were then processed using the GS, GA, and IGA methods to retrieve the phase image in each frame (since the object phase values are proportional to the illumination wavelengths, the employed algorithms were modified to account for this factor – see their implementation details in Supplementary Document 1 sections 1-4). The results in Fig. 6 are shown for the first of the collected frames. The time-lapse reconstructions of all frames for both datasets and each investigated algorithm are shown in Supplementary Videos 1-6.

The obtained reconstruction results confirm the ability of the IGA method to minimize twin image contribution with a quality similar to the GS method while maintaining the GA method's ability to average shot noise from multiple acquired images. It is noteworthy that the IGA method minimizes shot noise by averaging multiple frames (which are necessary to collect in multi-frame DIHM anyway), thus avoiding any loss of spatial resolution. This property is confirmed by the experiment, where the spermatozoid tails (approximately 0.4-0.5 µm width [57] – below the objective's diffraction-limited resolution of 0.8 µm) are visible with the same quality and sharpness in the IGA reconstruction as in the GS method. Moreover, in the GS LPB reconstruction, the sperm tail appears with worse contrast due to shot noise amplification. The IGA method averages the shot noise without blurring the tail, demonstrating its effectiveness in high-speed, low-photon imaging scenarios.

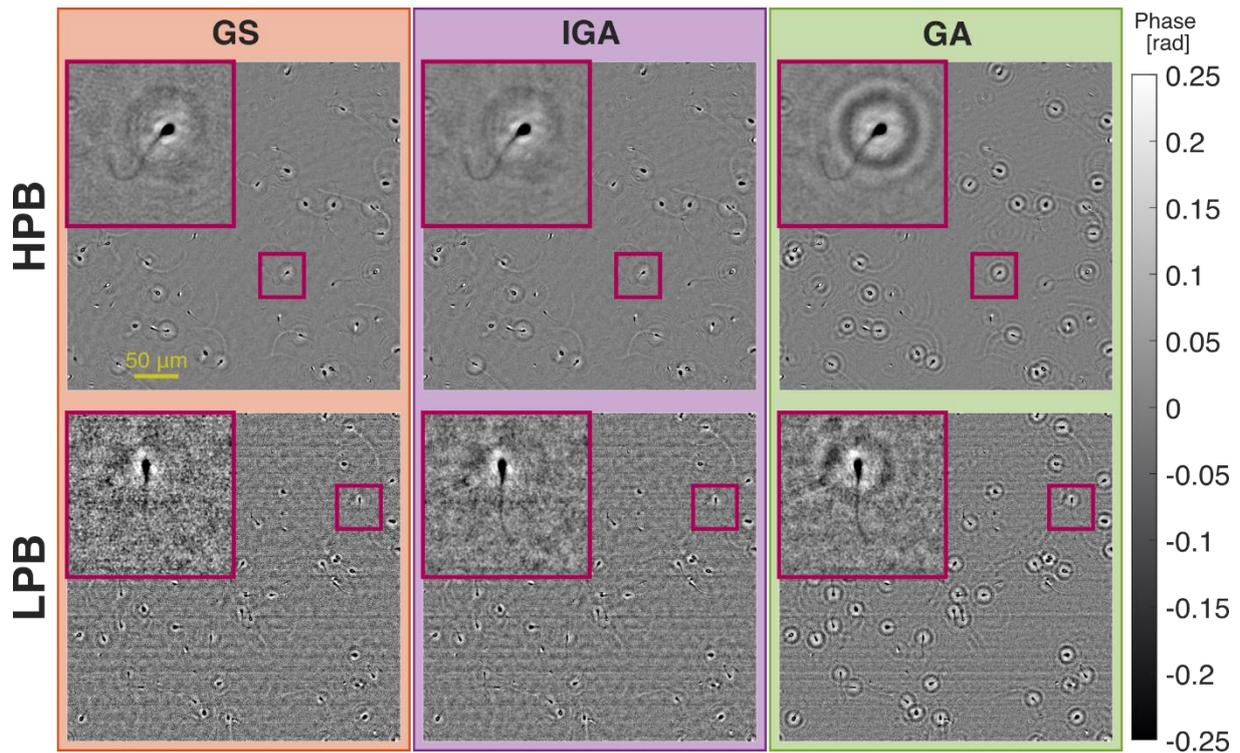

*Fig. 6. Phase images of a dynamic live human sperm sample reconstructed with investigated algorithms. Results are shown for first frames of HPB and LPB datasets. Full time-lapse reconstructions are shown in Supplementary Videos 1-6.*

The proposed IGA algorithm is not only limited to LPB imaging. It can be in general applied to any multi-frame DIHM system with low SNR data. This scenario may arise when imaging samples with extremely low optical thickness. Such objects scatter light minimally, resulting in holograms with low signal values (and therefore low SNR) even in HPB illumination regimes.

Figure 7 presents the QPI results of the lensless DIHM (Fig. 1(b)) of an optically thin phase resolution test target (custom-made - Lyncée Tec, Boroflat 33 glass). The test's physical thickness, given by the manufacturer is 16 ± 3 nm, corresponding to a phase change of 8.47 ± 1.59 [$10^{-2}$ rad] for the employed wavelength (561 nm). In the performed system, we collected three in-line holograms with different camera-sample distances (4.5, 6.5, 8.5 mm). The holograms were acquired with camera exposure times adjusted to the illumination intensity, resulting in a M value of 50.95 cgl – similar to HPB data in previous experiments. However, due to the weak scattering of light by the optically thin sample, the ΔM value was only 4.22 cgl, which is more comparable to LPB than HPB data from previous experiments.

The obtained results corroborate the usefulness of the proposed algorithm for DIHM reconstruction of low SNR data. The GA method suffers from twin image disturbances, visible mainly as a "halo" around larger test elements and as a diffraction pattern around optically thicker dust particles contaminating the sample. The GS algorithm successfully minimizes the twin image contribution, however, it amplifies the shot noise, which reduces the spatial resolution (compare zoomed-in elements from group Q in Fig. 7). The IGA method minimizes both noise types, as confirmed both qualitatively and quantitatively by calculating the S measurement from the object-free area marked with a yellow rectangle in Fig. 7. The height of the phase test elements estimated by the proposed method closely matches the reference height provided by the manufacturer (see the cross-section in Fig. 7). This demonstrates the IGA algorithm's effectiveness in accurately reconstructing low SNR holographic data, maintaining spatial resolution while minimizing both shot noise and twin image disturbances.

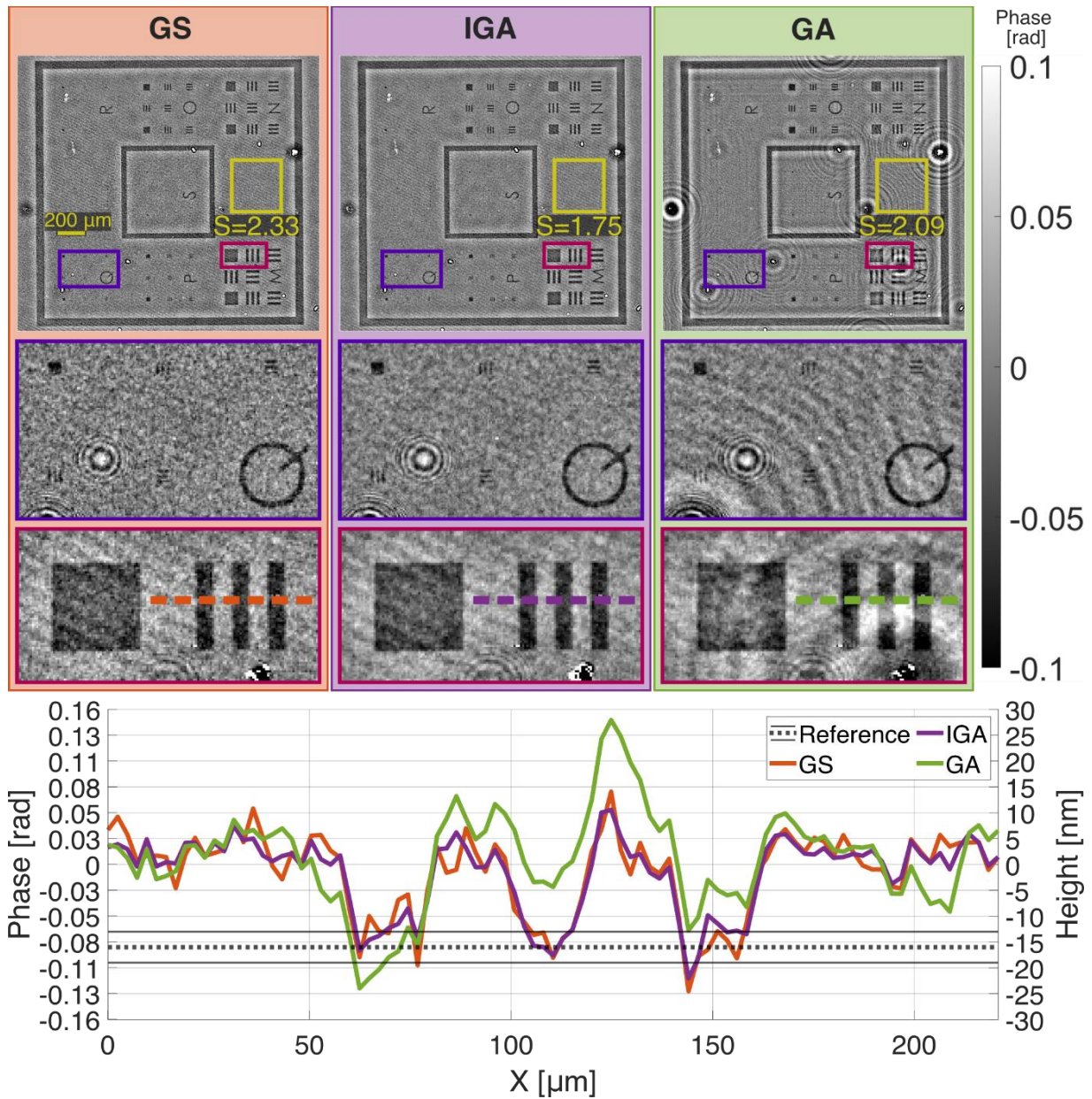

*Fig. 7. Top – phase reconstruction of optically thin resolution phase test target. S values are given in [10$^{-2}$ rad] units. Bottom – cross-section of the vertical elements from group M compared to the reference height measurement from white light interferometer.*

# Conclusion and Discussion

Low photon budget (LPB) imaging is crucial in various applications, particularly in photostimulation-free living cell imaging, high-speed imaging of dynamic biospecimens and in scenarios where maintaining the natural environment of live cell cultures is essential. The primary challenge in LPB imaging is to obtain high-quality images despite the reduced illumination, which leads to low SNR data.

As shown by the previous works [33], DIHM technique shows promising prospects in QPI under LPB conditions. However, the conventional DIHM reconstruction algorithms struggle in simultaneous minimizing of camera shot noise (present due to LPB) and twin image effect (inherent in DIHM). In this work, we addressed these challenges by proposing a novel IGA algorithm tailored for multi-frame DIHM under LPB conditions. The IGA algorithm combines the strengths of iterative phase retrieval methods and frame averaging to effectively minimize both twin image contribution and shot noise, what was validated numerically and experimentally. Additionally, the IGA algorithm proved effectiveness in reconstructing other kinds of low SNR DIHM data – HPB images of objects with extremely low optical thickness.

In summary, the proposed IGA algorithm offers a novel powerful tool for DIHM systems operating under LPB conditions, providing accurate phase reconstructions with minimized noise. This advancement opens new possibilities for high-speed imaging of dynamic biological samples and imaging of samples with extremely low optical thickness. Future work could explore the further optimization and automatization of parameters for specific imaging scenarios and the application of IGA algorithm in other computational imaging systems. In particular, the IGA based algorithms should be possible to implement in other QPI systems, where the iterative phase retrieval algorithms from multiplexed data are employed (e.g., ptychography [58] or Fourier ptychographic microscopy [59]).

# Funding

Funded by the European Union (ERC, NaNoLens, Project 101117392). Views and opinions expressed are however those of the author(s) only and do not necessarily reflect those of the European Union or the European Research Council Executive Agency (ERCEA). Neither the European Union nor the granting authority can be held responsible for them.

Research was funded by the Warsaw University of Technology within the Excellence Initiative: Research University (IDUB) programme (YOUNG PW 504/04496/1143/45.010008 and 504/04496/1143/45.010009).

M.R. is supported by the Foundation for Polish Science (FNP start programme).

This research has been funded by the Grant PID2020-120056 GB-C21 funded by MCIN/AEI/10.13039/501100011033. J. A. Picazo-Bueno is supported by the Spanish grant "Margarita Salas" (Ref. MS21-100), proposed by the Ministry of Universities of the Government of Spain (UP2021-044) funded by the European Union, NextGenerationEU.

# Data Availability Statement

In-line holograms employed in Figs. 5-7 are available at Ref. [60]. Employed IGA, GS and GA codes are available at Ref. [40]. Any other data may be obtained from the authors upon reasonable request.